\newcount\drawps\drawps=1 

\ifnum\drawps=1
\documentstyle[aps,twocolumn,prb,psfig]{revtex}
\else
 \documentstyle[preprint,aps,prb]{revtex}
\fi

\begin{document}
\draft
\title{
Low Temperature Static and Dynamic Behavior of the
 Two-Dimensional Easy Axis Heisenberg Model
}
\author{
M.\ E.\ Gouv\^ea,  G.M. Wysin,\cite{address1}  
S. A.  Leonel,
A.\ S.\ T.\ Pires, T. Kamppeter,\cite{address2} and  
F.G. Mertens\cite{address2}
}

\date{August 10, 1998}
\address{
Departamento de F\'{\i}sica, ICEx,
Universidade Federal de Minas Gerais 
Belo Horizonte, \\CP 702, CEP 30123-970, MG, Brazil
}

\maketitle
\ifnum\drawps=1
 \widetext
 \vskip-0.4in\hskip0.75in
 \parbox[t]{5.5in}{
\else
 \begin{abstract}
\fi
We apply the self-consistent harmonic approximation (SCHA) to study
static and dynamic properties of the two-dimensional classical
Heisenberg model with easy-axis anisotropy. 
The static properties
obtained are magnetization and spin wave energy as functions of
temperature, and the critical temperature as a function of the
easy-axis anisotropy. 
We also calculate the dynamic correlation 
functions using the SCHA renormalized spin wave energy.
Our analytical results, for both static properties
and dynamic correlation functions, are compared to
numerical simulation data combining cluster-Monte Carlo
algorithms and Spin Dynamics. 
The comparison allows us to 
conclude that far below the transition temperature, where the SCHA
is valid, spin waves are responsible for all relevant features 
observed in the numerical simulation data; topological 
excitations do not seem to contribute appreciably.
For temperatures closer to the transition temperature,
there are differences between the dynamic correlation 
functions from SCHA theory and Spin Dynamics; these
may be due to the presence of domain walls and solitons.

\ifnum\drawps=1
 \vskip0.07in
 \noindent
 PACS numbers: 75.10.Hk; 75.40.Cx }
 \narrowtext
\else
 \end{abstract}
 \pacs{ PACS numbers: 75.10.Hk; 75.40.Cx }
\fi

\section{INTRODUCTION}
\label{Intro}
Low-dimensional magnets have been extensively investigated
by many theorists and experimentalists in the last three decades.
More recently, 
the interest on the properties of two-dimensional(2D) 
Heisenberg magnets has been greatly revived since the 
discovery of high-T$_c$ superconductivity:
it is now well known\cite{Jerome}
 that the undoped, insulating La$_2$CuO$_4$ 
has a quasi-two-dimensional antiferromagnetic behavior.
However, most  quasi-two-dimensional magnetic real materials 
exhibit some kind of anisotropy: the anisotropic properties
often arise not so much from an anisotropy in the interaction
mechanism (which can be wholly isotropic) but from other
sources, such as the presence of a crystal field that couples 
the spins to a certain direction in the crystal.
Then, at least from a theoretical point of view,
 a  large amount of magnetic materials fits (under certain
circumstances like temperature range) into one of the two 
groups: easy-plane or easy-axis models.
Easy-plane 2D magnets have deserved a lot of attention
due to their possibility of showing
the topological Kosterlitz-Thouless phase transition.\cite{Kosterlitz}
The interest devoted to easy-axis magnetic systems has been
considerably smaller, specially concerning the study of its
dynamical properties.
It is our aim to address to this topic in this paper.

It must be emphasized that, although we shall be
concerned only with magnetic systems in this paper, many
of the magnetic Hamiltonians also allow for an 
interpretation other than a magnetic one.
Most physical problems concerning mutually interacting elements
that form a spatial array can be mapped into a magnetic Hamiltonian
\ifnum\drawps=1
\vbox to 196pt{\noindent }
\fi
by describing it within a pseudo spin formalism. 
The advantage of studying a general physical problem in its 
magnetic form is clearly that in magnetism several experimental
techniques are available to study the fundamental properties of
a system.\cite{Jongh}

The analysis of the general Ising-Heisenberg model is of
interest because, from the experimental point of view, the presence 
of some degree of anisotropy in the interaction is to be expected 
in nearly all cases.  
In addition, recently there has been a growing
interest in the study of topological excitations in the classical
two-dimensional easy-axis model.\cite{Ivanov}
Having finite excitation energy, the population of topological objects 
should be quite small at low temperatures.
Therefore, before taking into account the effect of topological 
excitations (solitons or similar objects) on the thermodynamics and 
dynamics of a system, we should consider the contribution of 
anisotropic spin waves. 
So we might ask:
can spin waves explain  experimental data or, in the absence
of experiments, computer simulation data at low temperatures?
This is the spirit and aim of this paper.
%

Here we consider the classical Heisenberg ferromagnet in two dimensions
(2D) with easy-axis exchange anisotropy
\begin{equation}\label{Hamiltonian}
H = - J 
\sum_{\bf{n},\bf{a}} {\bf{S}}_{\bf{n}} \cdot {\bf{S}}_{\bf{n}+\bf{a}}
- K \sum_{\bf{n},\bf{a}} S^z_{\bf{n}} S^z_{\bf{n}+\bf{a}} ~~~
\end{equation}
where the summations run over all distinct couples of spin sites 
$\bf{n}$ and its nearest neighbors $\bf{a}$. 
As the anisotropy parameter $K$ ranges from $0$ to $\infty$, we go 
from the isotropic 
%
%
\noindent
Heisenberg model to an Ising {\em{like}} model in which 
the spins tend to be confined along the $\pm z-$direction.
However, the resemblance to the Ising behavior can only hold
for  $T\ll K$: we find that, for Hamiltonian (\ref{Hamiltonian}),
$T_c \approx K$ for large $K$. 
This contrasts with $T_c\approx 2.27K$ for the 2D single-component
Ising model.

In addition to the usual domain walls we expect that there can be 
localized soliton-like excitations that can connect a 
small circular domain of positive $S^z$ to a 
surrounding region with negative $S^z$.
A spatial ``width'' of these objects (bubbles or droplets) can be 
estimated as approximately $\sqrt{J/K}$.
For intermediate values of $\sqrt{J/K}$, i.e., between a lattice constant
and the system size, these excitations can be important on a finite
discrete system.
These objects can also have a topological charge or winding number
of the spin field.
There was some indication in earlier Monte Carlo (MC)
simulations\cite{Costa94}
that they may play a role in the phase transition in this model;
their density was found to increase strongly passing through the 
transition temperature.
However, in a continuum static description they are found to be 
energetically unstable, according to the Derrick-Hobart 
theorem.\cite{Der-Hob}
Thus it makes sense to investigate whether it is necessary to
be aware of their presence in static and dynamic properties
of this model, or whether a description based on anisotropic
spin waves is sufficient. 

To this end we study the low temperature thermodynamics 
and dynamics of this model using a self consistent harmonic 
approximation theory (SCHA) to treat spin waves. 
As is well known, the SCHA is a reasonable approximation to calculate
the transition temperature and low temperature ($T<T_c$) properties of 
a system but it is of limited value in estimating critical properties.
Therefore, in our work, we did not attempt to do any calculation
for critical exponents and related aspects of a phase transition.
We compare the predictions of SCHA theory to numerical
simulations on  several $L \times L$ square lattices ($L=16,32,64,128$)
using Monte-Carlo and spin-dynamics (SD) simulations, which 
include effects due to all thermodynamically allowed excitations.
We present the thermodynamic results in Section \ref{Static}, and
in Section \ref{Dynamic}, the  calculation of the dynamical 
correlation function.
The simulation procedures are discussed in Sections \ref{MonteCarlo}
and \ref{Simulate}, and their comparison with the SCHA theoretical 
calculations is given in Section \ref{Results}. 
Finally, our conclusions are given in Section \ref{Conclude}.
 
\section{Static Properties}
\label{Static}

\subsection{Self-Consistent Harmonic Approximation}
Since its original derivation by Bloch,\cite{Bloch} the self
consistent harmonic approximation has been found to account for the low
temperature dependence of various properties of several magnetic
insulators, which seem to be fairly well-described by the Heisenberg 
model.\cite{Loly,Rastelli,Poling}
Its usefulness stems mainly from the way it takes into account a
substantial part of the interactions among spin waves, being 
characterized by simple temperature-dependent renormalization
factors for the unperturbed spin wave energy.

We start by writing the spin components using the Dyson-Maleev 
representation of spin operators
\begin{eqnarray}\label{transform}
S^x_{\bf n} &=& \frac{1}{2} 
\left[ \sqrt{2S} (a^{\dag}_{\bf n} + a_{\bf n}) \right] -
\frac{1}{\sqrt{8S}} a^{\dag}_{\bf n} a_{\bf n} a_{\bf n}
\nonumber \\
S^y_{\bf n} &=& \frac{1}{2i} 
\left[ \sqrt{2S} (a^{\dag}_{\bf n} - a_{\bf n}) \right] -
\frac{1}{\sqrt{8S}} a^{\dag}_{\bf n} a_{\bf n} a_{\bf n}
\\
S^z_{\bf n} &=& S - a^{\dag}_{\bf n}a_{\bf n}
\nonumber 
\end{eqnarray}
where $a^{\dag}_{\bf n}$ and $a_{\bf n}$ are the Bose spin operators
on site ${\bf n}$.
The harmonic spin wave Hamiltonian obtained from
(\ref{Hamiltonian}) is given by
\begin{equation}
H_0 = \sum_{\bf q} \omega_{\bf q} a^{\dag}_{\bf q} a_{\bf q}
\end{equation}
 where $a^{\dag}_{\bf q}$ and $a_{\bf q}$ are the Fourier transforms
of $a^{\dag}_{\bf n}$ and $a_{\bf n}$ respectively, and
\begin{equation}
\omega_{\bf q} = 4JS [ 1 - \gamma ({\bf q}) ] + 4KS
\end{equation}
with
$\gamma({\bf q}) = {1 \over 2} [\cos q_x + \cos q_y]$.
The spin wave approximation will be reasonable when
$\langle a^{\dag}_{\bf n}a_{\bf n}\rangle \ll S$,
so it ought to be fairly good for anisotropies 
satisfying the relation $T \ll 4 K S^2$.

Now we  simplify the general model by reducing Hamiltonian
(\ref{Hamiltonian}) to an effective harmonic problem with the 
effect of anharmonicity embodied in temperature-dependent
renormalized parameters.
This means that the couplings of the model are replaced
by quadratic interactions whose strength is then optimized.
Details of this method may be found in the 
literature\cite{Bloch,Rastelli} and we give here only
an outline of those steps pertinent to our present
calculation.

\begin{figure}
\psfig{file=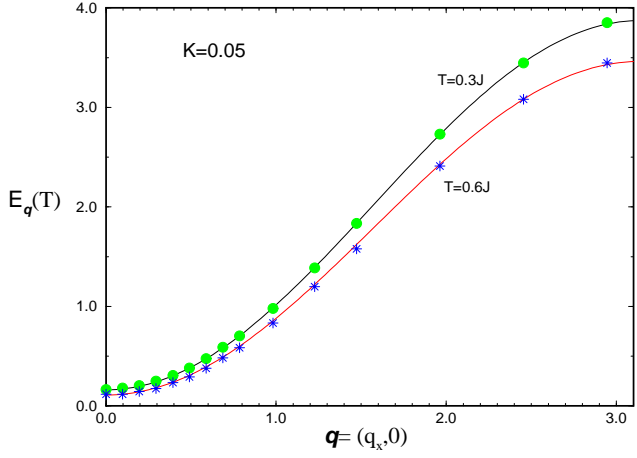,width=3.5in,angle=-90.0}
\caption{ \label{spinw} The curves 
correspond to the spin wave energy (from (\protect\ref{EdeT})) for
$T=0.3J$ (continuous), and $T=0.6J$ (dashed) for $K=0.05J$.
The circles and stars correspond to the values extracted from our 
numerical simulations; error bars are smaller than the symbols. }
\end{figure}

We assume as effective Hamiltonian the appropriate form for
a noninteracting gas of Bose excitations
\begin{equation}\label{Heffective}
\tilde{H}_0 = \sum_{\bf q} E_{\bf q} a^{\dag}_{\bf q} a_{\bf q}~~~.
\end{equation}
The spin wave energy is obtained by a variational procedure
based on the inequality for the Free energy $F$
\begin{equation}
F \le \tilde{F}_0 + \langle H - \tilde{H}_0 \rangle_0~~~,
\end{equation}
where the brackets indicate the thermal average.
Traces should be taken only over the physical states, that is,
states with no more than $2S$ spin deviations on a single site.
The minimization of (\ref{Heffective}) with respect to $E_{\bf q}$
determines the spin wave energies. We obtain, in the classical
limit, following Rastelli et al,\cite{Rastelli} 
\begin{eqnarray}\label{EdeT}
E_{\bf q} (T) &=& 4 J S \left( 1 - \gamma ({\bf q}) \right)
\left[ 1 - \tilde{\beta} (T) + \tilde{\eta}(T) \right]
\nonumber \\
&+& 4 K S \left[ 1 - \tilde{\beta}(T) - \gamma (\bf q)
\tilde{\eta}(T) \right]
\end{eqnarray}
where
\begin{equation}\label{beta}
\tilde{\beta} (T) = \frac{T}{NS} \sum_{\bf q} \frac{1}{E_{\bf q}}
~~,
\end{equation}
\begin{equation}\label{eta}
\tilde{\eta} (T) = 
\frac{T}{NS} \sum_{\bf q} \frac{\gamma(\bf q)}{E_{\bf q}}
~~,
\end{equation}
where $N$ is the number of sites.
Eqs. (\ref{EdeT}), (\ref{beta}), and (\ref{eta}) are coupled equations 
which we solved self consistently by an iterative method.
These coupled equations have a double-valued  
solution below $T_c$ and no real solution above $T_c$: 
this is the typical behavior for self consistent harmonic 
approximations and allows for easy determination of $T_c$.
The lower branch (for $T<T_c$) has an unphysical
temperature dependence and may be discarded as a spurious 
mathematical solution that is physically unstable.
In Figure \ref{spinw} the spin wave energy for $K/J=0.05$ is given for 
two temperatures well below $T_c\approx 0.75J$: $T=0.3J$, and  
$T=0.6J$. 
The circles and stars shown in Figure \ref{spinw} were 
taken from our numerical
simulation data (to be described in Section \ref{Simulate}).
As can be seen, the comparison between the SCHA and numerical 
results is remarkably good: the SCHA describes well the decrease
of energy with increasing temperature (and, also, the energy 
dependence with the wavevector).

\begin{figure}
\psfig{file=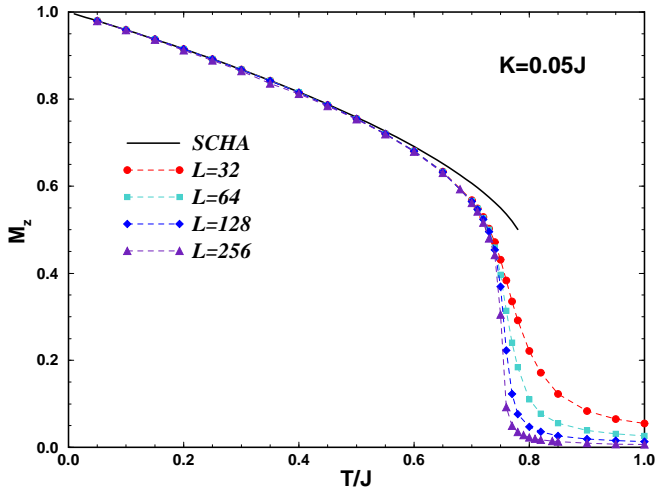,width=3.5in,angle=-90.0}
\caption{ \label{Mz05} Magnetization as a function of
temperature for $K=0.05J$. Solid curve is the SCHA theory.
Various symbols correspond to MC simulation for indicated
system sizes; error bars are smaller than the symbols.}
\end{figure}

The reduced spontaneous magnetization along the $z$-axis is 
given by
\begin{equation}\label{magn}
\frac{M_z(T)}{M_z(0)} = 1 - \tilde{\beta} (T)~~~.
\end{equation}
In Figure \ref{Mz05}, we present results obtained from 
Eq. (\ref{magn}) for $K/J=0.05$ and
compare to our Monte Carlo (MC) data (obtained as described
in Section \ref{MonteCarlo}).
The slight overestimate of $T_c$ from SCHA clearly is due to 
the fact that it does not include all possible modes of
fluctuations, that are included in the MC calculations.
The SCHA theory has no built-in requirement to make
the magnetization null at the critical temperature $T_c$, and 
consequently, we find, as Figure \ref{Mz05} shows, a nonzero 
value for $M_z(T_c)$. 
This is  typical of SCHA approaches: the theory applies only
below $T_c$, and for temperatures $T \ge T_c$ the magnetization
is taken as zero, implying a discontinuous jump at $T_c$.
In fact, the scaling of the MC data for $M_z(T)$ with system size $L$
strongly suggests the presence of a discontinuous jump.
%

\subsection{Monte Carlo}
\label{MonteCarlo}
In order to evaluate the accuracy of the above theory, we 
calculated $T_c$ and the magnetization and other thermodynamic 
quantities using a hybrid classical Monte Carlo approach
on periodic $L\times L$ square lattices. 
We applied a combination of Metropolis single-spin moves and
over-relaxation moves that modify all three spin components, 
and in addition, Wolff single-cluster operations\cite{Wolff}
 that modify {\em only} the $S^z$ components.  
The over-relaxation and cluster moves are necessary to avoid 
critical slowing down near $T_c$, which is tending to freeze the 
$S^z$ components.
The single spin and over-relaxation moves are standard, here we
give only a few details about the cluster algorithm.
In the Wolff single-cluster algorithm, the cluster-flip operation 
we used only reverses the sign of $S^z$ for all sites that have been 
included into the cluster.
This is reminiscent of the Swendsen-Wang algorithm\cite{SW} for 
Ising models, but we only build one cluster at a time as in the 
Wolff algorithm.
The cluster moves cannot be used alone because they do not
change the {\em magnitudes} of $S^z$ spin components.

A cluster is built up starting from a randomly chosen
seed site ${\bf n}$, immediately inverting its $S^z$ component: 
$S_{\bf n}^z \rightarrow -S_{\bf n}^z$, and then including neighboring 
sites ${\bf n+a}$ with a probability,
\begin{equation}
\label{pbond}
p_{\rm bond} = {\rm max}
	\left[0, 1-e^{-\beta \Delta E_{{\bf n},{\bf n+a}} }\right].
\end{equation}
Here $\Delta E_{{\bf n},{\bf n+a}}$ is the energy change involved
if site ${\bf n+a}$ is {\em not} flipped:
\begin{equation}
\label{dE}
\Delta E_{{\bf n},{\bf n+a}} =  2 (J+K) S_{\bf n}^{z} S_{\bf n+a}^{z}.
\end{equation}
Note that in this formula site ${\bf n}$  was already included into
the cluster and $S_{\bf n}^{z}$ was already inverted.
Eq.\ (\ref{pbond}) represents the cluster growth as essentially
a sequence of Metropolis decisions, according to whether
$\Delta E_{{\bf n},{\bf n+a}}$ is less than or greater than zero.   
Newly included sites then have their neighbors tested for inclusion
until the cluster is done growing, at which point all included
sites have already been modified.

We define one cluster sweep as building enough single clusters until
the number of sites included into clusters is one quarter of the total 
number of sites in the system.  
Then we defined one hybrid Monte Carlo step as one over-relaxation
sweep followed by one Metropolis single spin sweep followed by
one Wolff cluster sweep.
Equilibrium data shown here are averages over $10^5$ to $4\times 10^5$
Monte Carlo steps.
The critical temperature was determined from the change in the 
distribution of $z$-component of total magnetization, which is easily
characterized by Binder's fourth cumulant ratio,\cite{Binder}
\begin{equation}
\label{Ulz}
U_L = 1 - { {\langle M_z^4 \rangle} \over 
	    {3 \langle M_z^2 \rangle ^2} }.
\end{equation}
The crossing point of curves of $U_L(T)$ for different system sizes   
gives a good estimate of $T_c$.  
All calculations were made for square lattices of size $L\times L$,
using unit spins $S=1$ and fixing $J=1$ while allowing $K$ to be 
varied.
%

\subsection{Static Results}
\begin{figure}
\psfig{file=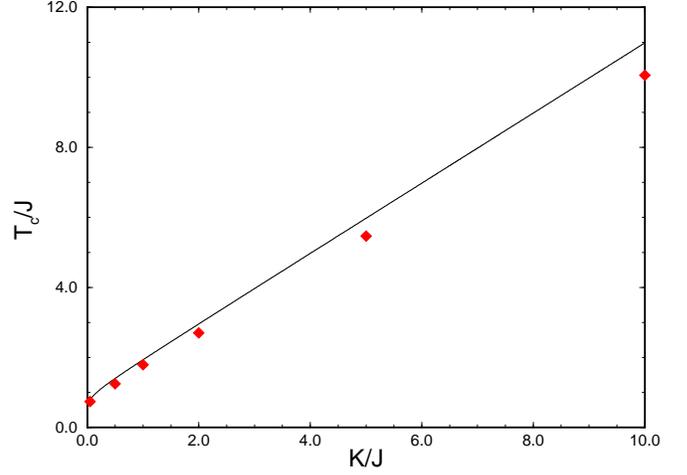,width=3.5in,angle=-90.0}
\caption{ \label{tempcr} Critical temperature as a function
of the anisotropy parameter $K/J$. 
The symbols correspond to
the values obtained in our MC calculation (as described in
Section \protect\ref{MonteCarlo}); error bars are smaller than the symbols. }
\end{figure}

The critical temperature from the SCHA as a function of anisotropy 
parameter $K/J$ is shown in Figure \ref{tempcr} and compared with 
numerical MC estimates for a set of specific values of $K/J$ ranging 
from $0.05$ to $10.0$.
Generally, the SCHA overestimates $T_c$ when compared to the
MC results, because it does not fully take into account all
possible fluctuations that are taking part in the transition.
Notice that, as $K$ increases, the dependence of $T_c$ on $K/J$
becomes linear.
For $K/J \gg 1$, we recover a continuous spin Ising Hamiltonian:  
Eq.\ (\ref{Hamiltonian}) can be approximated as
$H \approx J(1 + K/J)S^z_{\bf n} S^z_{{\bf n}+{\bf a}} =
\tilde{K} S^z_{\bf n} S^z_{{\bf n}+{\bf a}}$.
Figure \ref{tempcr}  shows that, for $K/J > 1.0$, the results follow a 
straight line with slope $\approx 1.0$.
We remark again that, strictly speaking, the analogy between
Hamiltonian (\ref{Hamiltonian}) and the continuous spin 
Ising model can only be expected to hold for $T\ll K$.
For moderate and high temperatures, model (\ref{Hamiltonian})
still exhibits the full entropy effects of a three-component
spin model, resulting in a much lower transition temperature
than a one-component Ising model. 
For this reason we cannot expect to compare our results to the 
ones obtained for the usual 2D Ising model.

Some of the drawbacks of the SCHA are well known:
(i) it does not take into account strong coupling effects
which are important at high temperatures and at short
wavelengths; (ii) it also neglects the kinematical 
interaction and gives a first-order phase transition 
to the paramagnetic phase (where the true phase transition 
should be of second order). 
Notwithstanding this, we
see that the theory gives results which compare quite
well with the MC data we obtained.
%

This good agreement cannot be used to conclude that solitons
do not have an important contribution to the properties of
our model. 
As is well known, in the one dimensional easy-axis ferromagnet,
the soliton connects two distinct ground states and has, therefore,
a {\em global} effect in the system.\cite{beth}
As a consequence, a pure spin wave calculation does not predict
correctly all thermodynamic quantities.
For instance, spin waves give a linear behavior with
temperature $T$ (for $T\rightarrow 0$) for the correlation
length, while the soliton model predicts correctly an
exponential behavior.
In two dimensions,  however, the soliton has only a {\em local}
effect and its contribution to thermodynamic quantities
should be small.
The reasonable agreement of the SCHA calculation with
the Monte Carlo data is not, per se,
an indication that we should rule out topological effects.
The signature of topological solitons is best analyzed 
in the dynamics where it should manifest as a central peak.
This topic will be discussed in the following Sections.

\section{Dynamic Correlation Functions}
\label{Dynamic}

\subsection{SCHA}
\begin{figure}
\psfig{file=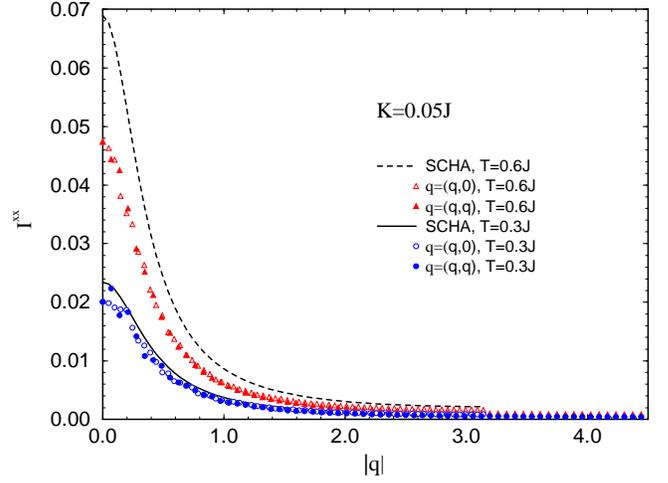,width=3.5in,angle=-90.0}
\caption{ \label{Ixx}  In-plane integrated intensity 
$I^{xx}$ versus wavevector, from SCHA (curves)
compared with MC-SD (symbols) for $K=0.05J$, 
$L=128$.}
\end{figure}

From Hamiltonian (\ref{Hamiltonian}) and using Eqs.\ (\ref{transform}),
we obtain the time dependent correlation functions
\begin{eqnarray}
\langle S^x_{\bf q}(t) S^x_{-\bf q} \rangle  &=&
\frac{S}{2} \langle [a_{\bf q}(t) + a_{\bf q}^{\dag}(t)]
[a_{-\bf q} + a_{-\bf q}^{\dag}] \rangle ~~~,
\nonumber \\
\langle S^y_{\bf q}(t) S^y_{-\bf q} \rangle  &=&
\frac{S}{2} \langle [a_{\bf q}(t) - a_{\bf q}^{\dag}(t)]
[a_{-\bf q} - a_{-\bf q}^{\dag}] \rangle  ~~~,
\nonumber \\
\langle S^z_{\bf q}(t) S^z_{-\bf q} \rangle  &=&
 \delta_{{\bf q},0} \langle S-( a^{\dag}_l a_l)\rangle ^2 
\\  &+&
\langle\delta S^z_{\bf q}(t)\delta S^z_{-\bf q}\rangle ~~~,
\nonumber
\end{eqnarray}
where 
\begin{equation}
\delta S^z_{\bf q}(t) = \frac{1}{N^{1/2}} 
\sum_{\bf l} e^{i {\bf q} \cdot {\bf l}} 
(a^{\dag}_{\bf l}(t) a_{\bf l}(t) 
- \langle a^{\dag}_{\bf l}(t) a_{\bf l}(t)\rangle ).
\end{equation}
The averages are readily evaluated, and give the
$xx$- and $yy$- dynamical correlation functions:
\begin{eqnarray}\label{Sxx}
\langle  S^x_{\bf q}(t) S^x_{-{\bf q}} \rangle  &=&
\langle  S^y_{\bf q}(t) S^y_{-{\bf q}} \rangle  
\\ &=&
\frac{S}{2} \left[ n_{\bf q} e^{i E_{\bf q} t} +
(n_{\bf q} + 1)  e^{-i E_{\bf q} t} \right]~~~,
\nonumber
\end{eqnarray}
where $ n_{\bf q} $  is the Bose occupation number and $E_{\bf q}$
is the self-consistent spinwave frequency of Eq.\ (\ref{Heffective}).
Eq.\ (\ref{Sxx}) leads to pure spin wave peaks for the spectral 
function:\cite{FTnote} 
\begin{equation} \label{Sxxqw}
S^{xx}({\bf q}, \omega) = \frac{S}{2(2\pi)^2}
\left[ n_{\bf q} \delta(\omega-E_{\bf q}) 
 + (n_{\bf q}+1) \delta(\omega+E_{\bf q}) \right].
\end{equation}
The comparison between these SCHA results for $S^{xx}$
and those obtained by spin dynamics simulation (Sec.\ \ref{Simulate})
are shown in Figures \ref{spinw} and \ref{Ixx}.
These are discussed in more detail in Sec.\ \ref{Results}.B below.

The $zz$- correlation contains, in addition to the Bragg scattering 
at ${\bf q}={\bf 0}$, a term describing correlations in the 
fluctuations of the spin's $z$ component.
Evaluating the averages, we obtain
\begin{eqnarray}\label{Sqt}
\langle \delta S^z_{\bf q}(t)\delta S^z_{-\bf q}\rangle  &=&
\frac{1}{2N} \sum_{\bf k}
\left[ e^{i \Omega t} 
\left( 1 + n_{\frac{{\bf q}}{2} - {\bf k}} \right) 
 n_{\frac{\bf q}{2} + {\bf k}}
\right. \nonumber \\
&+& \left.
 e^{-i \Omega t} 
\left( 1 + n_{\frac{\bf q}{2} + {\bf k}} \right)
n_{\frac{\bf q}{2} - {\bf k}}
\right]
\end{eqnarray}
where 
\begin{equation}\label{Omega}
\Omega=E_{\frac{\bf q}{2} + {\bf k}} -
E_{\frac{\bf q}{2} - {\bf k}}~~~. 
\end{equation}
Eq.\ (\ref{Sqt}) corresponds to the various two-spin-wave
scattering terms. 
It is interesting to notice that,  to this order, only {\em difference} 
processes contribute to the dynamics.
The time Fourier transform of (\ref{Sqt}), together with the 
thermodynamic limit $L\rightarrow \infty$, gives us the response 
function\cite{FTnote} 
\begin{eqnarray}\label{Szz}
S^{zz}({\bf q}, \omega) &=&
\frac{1}{2(2\pi)^4} \int d{\bf k} 
\left[
n_{\frac{\bf q}{2} + {\bf k}}
\left( 1 + n_{\frac{\bf q}{2} - {\bf k}} \right)
\delta (\omega + \Omega)
\right. 
\nonumber \\
 &+& \left.
n_{\frac{\bf q}{2} - {\bf k}}
\left( 1 + n_{\frac{\bf q}{2} + {\bf k}} \right)
\delta (\omega - \Omega)
\right].
\end{eqnarray}
Using the delta functions we obtain integrals on the contours
${\cal C}^{\pm}$ defined by $\omega=\pm \Omega$:  
the first integral is
\begin{equation}
\label{contour}
\frac{1}{2(2\pi)^4} \int_{{\cal C}^+} dl_{\bf k} n_{\frac{\bf q}{2} + {\bf k}}
\left( 1 + n_{\frac{\bf q}{2} - {\bf k}} \right) {\mid \nabla_k \Omega \mid}^{-1},
\end{equation}
where $dl_{\bf k}$ is the contour element and ${\mid \nabla_k \Omega \mid}$ 
designates the Jacobian of the involved transformation.
There is a singularity in these integrands for every
minimum or maximum of $\Omega$ and the spectrum is in general
quite complicated.
We emphasize that, here, we have also used the self-consistent result 
obtained in Section \ref{Static} for the spin wave energies.
%
%
Results for the spectral functions  obtained from (\ref{Szz}) 
will be compared with those from MC-SD simulations below in
Section \ref{Results}.

\subsection{Spin Dynamics Simulations}
\label{Simulate}
The spin dynamics simulation is 
standard.\cite{Kawabata+86,Wysin+90,Evertz}
Here we summarize the method and describe the particular numerical 
parameters used.
For a given temperature, a set of 200 initial states was taken from
the Monte Carlo simulation to serve as initial conditions for the
spin-dynamics time integration.
The nonlinear equations of motion associated with Hamiltonian 
(\ref{Hamiltonian}) are
\begin{equation}
{d \bf{S}_{\bf{n}} \over dt} = \bf{S}_{\bf{n}} \times 
   \left[\tilde{J} \sum_{\bf{a}} \bf{S}_{\bf{n}+\bf{a}} \right],
\end{equation}
where $\tilde{J}$ is the diagonal matrix of exchange couplings,
\begin{equation}
\tilde{J} =  
\left( \begin{array}{ccc}
           J   &  0  &  0     \\
           0   &  J  &  0     \\
           0   &  0  & J+K  
\end{array}  \right)
\end{equation}
These were integrated forward in time using a standard fourth order
Runge-Kutta scheme with time step $h=0.035/J$ (for small $K/J$).
By saving data for time Fourier transforms at intervals $dt=6h$,
allows for measuring $S({\bf q},\omega)$ out to $\omega_{\rm max}
=2\pi/dt \approx 30 J$.   
We saved a total of $N_t=2^{11}$ samples in time, integrating out to 
final time $t_{\rm max}=N_t dt \approx 430/J$, giving a frequency 
resolution of $\delta\omega = 2\pi/t_{\rm max} \approx 0.015J$. 
The  space and time Fourier-transformed spin-spin correlations
were averaged over the 200 initial states to get $S({\bf q},\omega)$,
for both in-plane and out-of-plane spin components.


\section{DYNAMIC CORRELATIONS: RESULTS}
\label{Results}

\subsection{Small Lattices ($L\le 64$)}
At low temperatures $T\ll T_c$, we especially expect that the SCHA 
should give good agreement with the spin-dynamics simulation.
We had noticed, however, that spin dynamics for small lattices 
gives an interesting set of unevenly spaced peaks in 
$S^{zz}({\bf q}, \omega)$, in contrast to one sharp peak at the
spinwave frequency in $S^{xx}({\bf q}, \omega)$,
and also in contrast to the smooth behavior predicted
for $S^{zz}({\bf q}, \omega)$ by Eq.\ (\ref{Szz}).
Also, an intensity {\em maximum} in $S^{zz}({\bf q}, \omega)$ for  
$\omega\rightarrow 0$ is present for wavevectors of the form
${\bf q} = {2\pi\over L} ( 2m, 0)$, where $m$ is an integer.
On the other hand, for wavevectors 
${\bf q} = {2\pi\over L} ( 2m+1, 0)$, there is an intensity 
{\em minimum} in $S^{zz}({\bf q}, \omega)$ at $\omega\rightarrow 0$. 
%
%
In order to see if the SCHA theory could explain this interesting
result we re-started our calculation from (\ref{Sqt}), restricting
the sums to the discrete wavevectors
${\bf k} = \frac{2\pi}{L} (m,n)$ of each lattice. 
Also, the time integration for $S({\bf q},\omega )$ was performed
for a {\em finite} time interval $t_{max}$, 
\begin{equation}\label{Discrete}
S^{zz}({\bf q},\omega) =
 \frac{1}{2\pi} \int_{-t_{max}/2}^{t_{max}/2} 
S^{zz}({\bf q},t) e^{-i \omega t} dt
\end{equation}
where $t_{max}$ was taken to be the same as the integration
time ($430/J$) used in our simulations.
%
Eq.\  (\ref{Sqt}) is modified for a finite time interval, and a 
complete analysis leads to
\begin{eqnarray}\label{disc}
S^{zz}&(&{\bf q},\omega) =
\frac{t_{max}}{2N(2\pi)^3} \sum_{\bf k} \nonumber \\
&& \left\{ 
n_{\frac{\bf q}{2} + {\bf k}}
\left( 1 + n_{\frac{\bf q}{2} - {\bf k}} \right)
\left[\frac{\sin [(\omega - \Omega)t_{max}/2]}
{(\omega - \Omega)t_{max}/2}\right]^2
\right. \nonumber \\
&&+ \left.
 n_{\frac{\bf q}{2} - {\bf k}}
\left( 1 + n_{\frac{\bf q}{2} + {\bf k}} \right)
\left[\frac{\sin [(\omega + \Omega)t_{max}/2]}
{(\omega + \Omega)t_{max}/2}\right]^2
\right\}.
\end{eqnarray}
The expression can be thought to represent $S({\bf q},\omega)$
as a sum over a set of narrow peaks of width approximately $2/t_{max}$,
centered at frequencies $\Omega$, determined by choosing ${\bf k}$ 
such that both $\frac{\bf q}{2} + {\bf k}$ and 
$\frac{\bf q}{2} - {\bf k}$ in Eq.\ (\ref{Omega}) are allowed discrete 
wavevectors.
Besides restricting the sum in (\ref{disc}) to the discrete
set of lattice wavevectors, the finite time integration $t_{max}$
implies discrete frequency increments 
$\delta \omega =2\pi/t_{max} \approx 0.015J$, 
the same as in our spin dynamics simulation.

\begin{figure}
\psfig{file=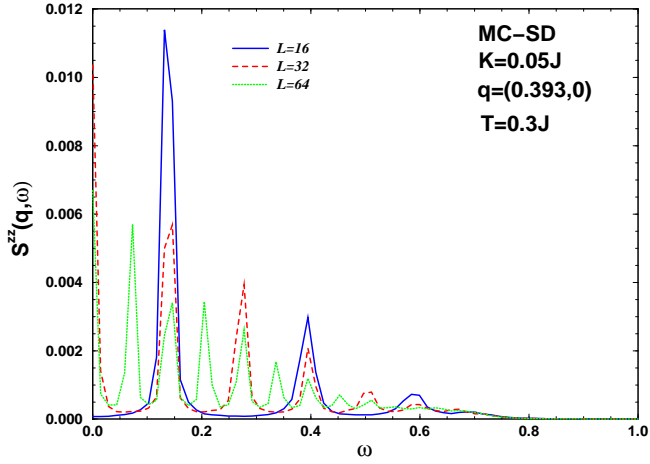,width=3.5in,angle=-90.0}
\caption{ \label{MC-Small} $S^{zz}({\bf q},\omega)$ obtained from
numerical simulation for $K=0.05J$, $T=0.3J$, and $L=16,32,64$, and
wavevector ${\bf q}=(0.393,0)$. }
\end{figure}

\begin{figure}
\psfig{file=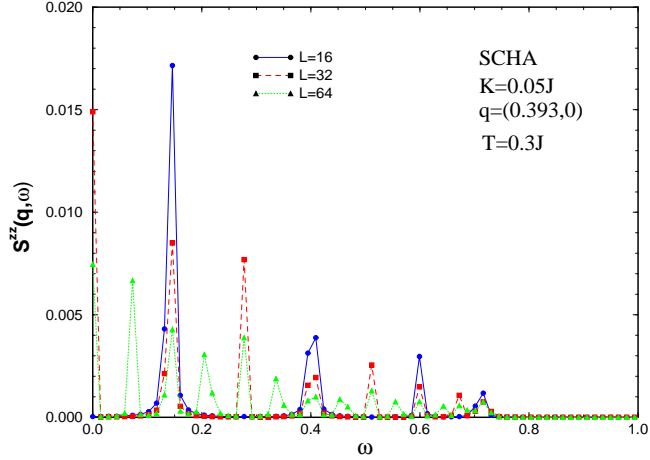,width=3.5in,angle=-90.0}
\caption{ \label{DS-Small}  $S^{zz}({\bf q},\omega)$ obtained from
discrete summation (\protect\ref{disc}) for $K=0.05J$, $T=0.3J$, and 
$L=16,32,64$, and wavevector ${\bf q}=(0.393,0)$. }
\end{figure}

Examination of (\ref{Omega}) and (\ref{disc})
allows us to conclude  that a nonzero intensity 
in $S^{zz}({\bf q},\omega\rightarrow 0)$ can exist for {\em all} wavevectors
and not only for those of the form ${\bf q}=\frac{2\pi}{L}(2m,0)$.
However, a little consideration shows that if ${\bf q/2}$
does not fall on a reciprocal lattice vector, then it is impossible
to choose a value of ${\bf k}$ in Eq.\ (\ref{Omega}) to
give $\Omega=0$. 
Therefore, for wavevectors ${\bf q} = {2\pi\over L} ( 2m+1, 0)$,  
none of the multiple peaks in (\ref{disc}) will be centered at 
zero frequency, and $S^{zz}({\bf q},\omega\rightarrow 0)$ is a 
local minimum. 
Although no peak is centered at zero, the tails can contribute there.
On the other hand, for wavevectors such as 
${\bf q} = {2\pi\over L} ( 2m, 0)$, and 
${\bf q} = {2\pi\over L} ( m, m)$, we see that $\frac{\bf q}{2}$ 
falls on a highly symmetric point in the reciprocal lattice, and 
it is always possible to choose ${\bf k}$ to get $\Omega=0$ in 
Eq.\ (\ref{Omega}).
Then for these cases, there is a peak  at zero frequency, and
$S^{zz}({\bf q},\omega\rightarrow 0)$ is a local maximum.
The overall behavior of $S^{zz}({\bf q},\omega)$ with
the lattice size obtained either by numerical simulation 
(Fig.\ \ref{MC-Small}) or by the calculation of Eq. (\ref{disc})
(Fig.\ \ref{DS-Small}) agree very well.
In order to make this comparison, because the spin-dynamics simulations
are purely classical, it is necessary to replace all factors of
$(1+n_{\bf q})$ in the SCHA expressions by $n_{\bf q}$.
Also, these occupation numbers were evaluated by their
classical limit, $n_{\bf q}=T/E_{\bf q}$, consistently with
the static calculations in Sec.\ \ref{Static}.
Figures \ref{MC-Small} and \ref{DS-Small} were obtained for
$K/J=0.05$ (where $T_c\approx 0.75 J$), ${\bf q}=(0.393,0)$, and
$T=0.3J$ for lattice sizes $L=16$, $32$, and $64$.

Comparing the several peaks shown in Figures \ref{MC-Small} and 
\ref{DS-Small}, for a specific value of $L$, we see that they are 
positioned around the same frequencies.
The important feature is that as the system size is increased,
the spacing between the multiple peaks in $S^{zz}({\bf q}, \omega)$
becomes smaller as ${1\over L}$.
In addition, for a longer time integration $t_{max}$, the widths
of the peaks will be narrower, and therefore they will become
more distinct. 
As far as we aware, this strong finite-size effect in low-temperature
spin-dynamics simulations is a feature that has been previously 
ignored.
It is very likely, however, that it appears in any related models.
For example, finite-size effects most likely explain similar 
low-temperature multiple-peak features that have appeared in 
$S({\bf q}, \omega)$ calculated for the 2D Heisenberg model with 
easy-plane anisotropy.\cite{Evertz,GouveaWysin}

\subsection{Large Lattices ($L\ge 128$) }
The SCHA calculation [Eq.\ (\ref{Sxxqw})] and the
MC-SD simulations both give single narrow spinwave peaks in 
$S^{xx}({\bf q}, \omega)$, regardless of lattice size.
The MC-SD peak positions for $L=128$ have been compared with the
SHCA results in Fig.\ \ref{spinw}, and agree very well for
the temperatures studied.
The SCHA theory gives peaks of zero width, thus it makes sense
to compare the integrated intensities for the positive
frequency peak, $I^{xx}=\int_{0}^{+\infty} dw S^{xx}({\bf q}, \omega)$.
These are shown in Fig.\ \ref{Ixx}, where the MC-SD results are
compared to those obtained from Eq.\ \ref{Sxxqw},
$I^{xx}=S n_{\bf q}/(8\pi^2)$.
For the lower temperature, $T=0.3J$, there is very good agreement.  
The good low-T agreement, with no adjusted parameters, shows
that the approximations made in the SHCA theory are reasonable
where we expect this simple theory to work.
For $T=0.6J$, however, the MC-SD result is suppressed compared 
to the SCHA prediction.
Currently we cannot say whether this suppression should be
better described by spinwave interaction terms or possibly
by nonlinear excitations such as solitons or domain walls.
Clearly, both effects could become more important as
the critical temperature is approached.

\begin{figure}
\psfig{file=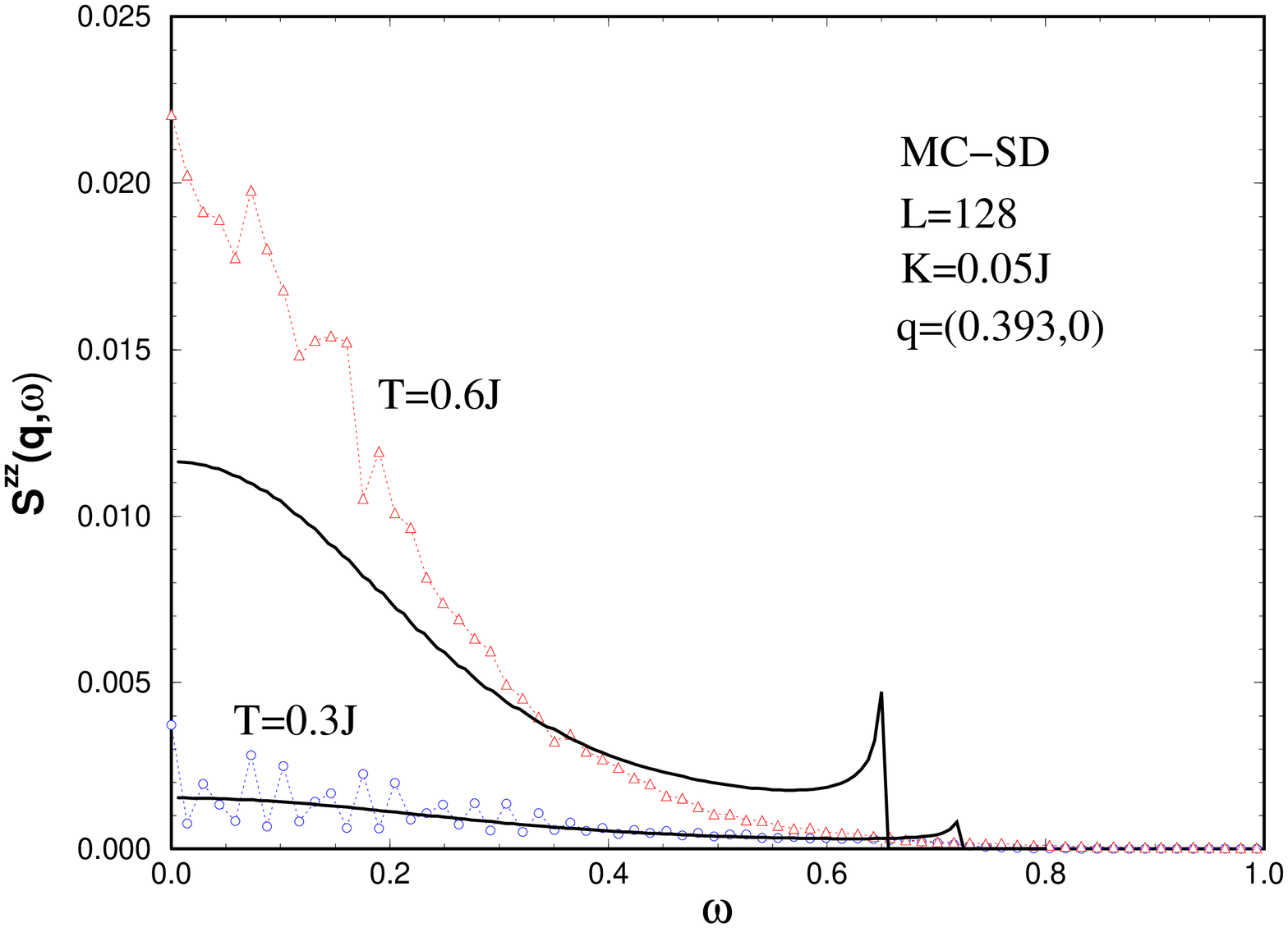,width=3.5in,angle=-90.0}
\caption{ \label{Szz0393}  $S^{zz}({\bf q},\omega)$ from
(continuous line) continuum limit (\protect\ref{Szz})  and from (circles and 
triangles) numerical simulation for $K=0.05J$, $T=0.3J$, and $T=0.6J$, and
wavevector ${\bf q}=(0.393,0)$.}
\end{figure}

\begin{figure}
\psfig{file=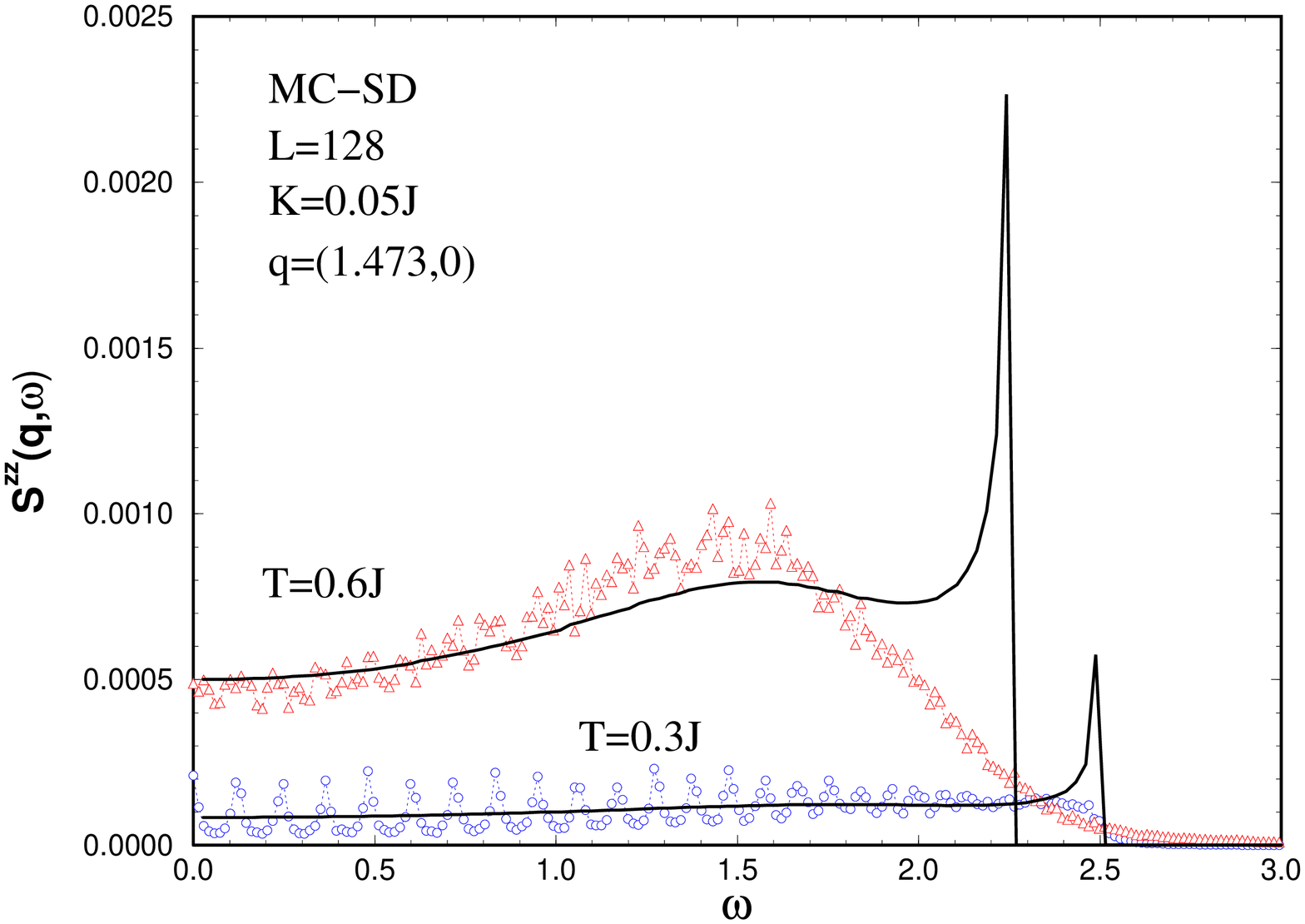,width=3.5in,angle=-90.0}
\caption{ \label{Szz1473}  $S^{zz}({\bf q},\omega)$ from
(continuous line) continuum limit (\protect\ref{Szz})  and from (circles and
triangles) numerical simulation
for $K=0.05J$, $T=0.3J$, and $T=0.6J$, and
wavevector ${\bf q}=(1.473,0)$.}
\end{figure}

\begin{figure}
\psfig{file=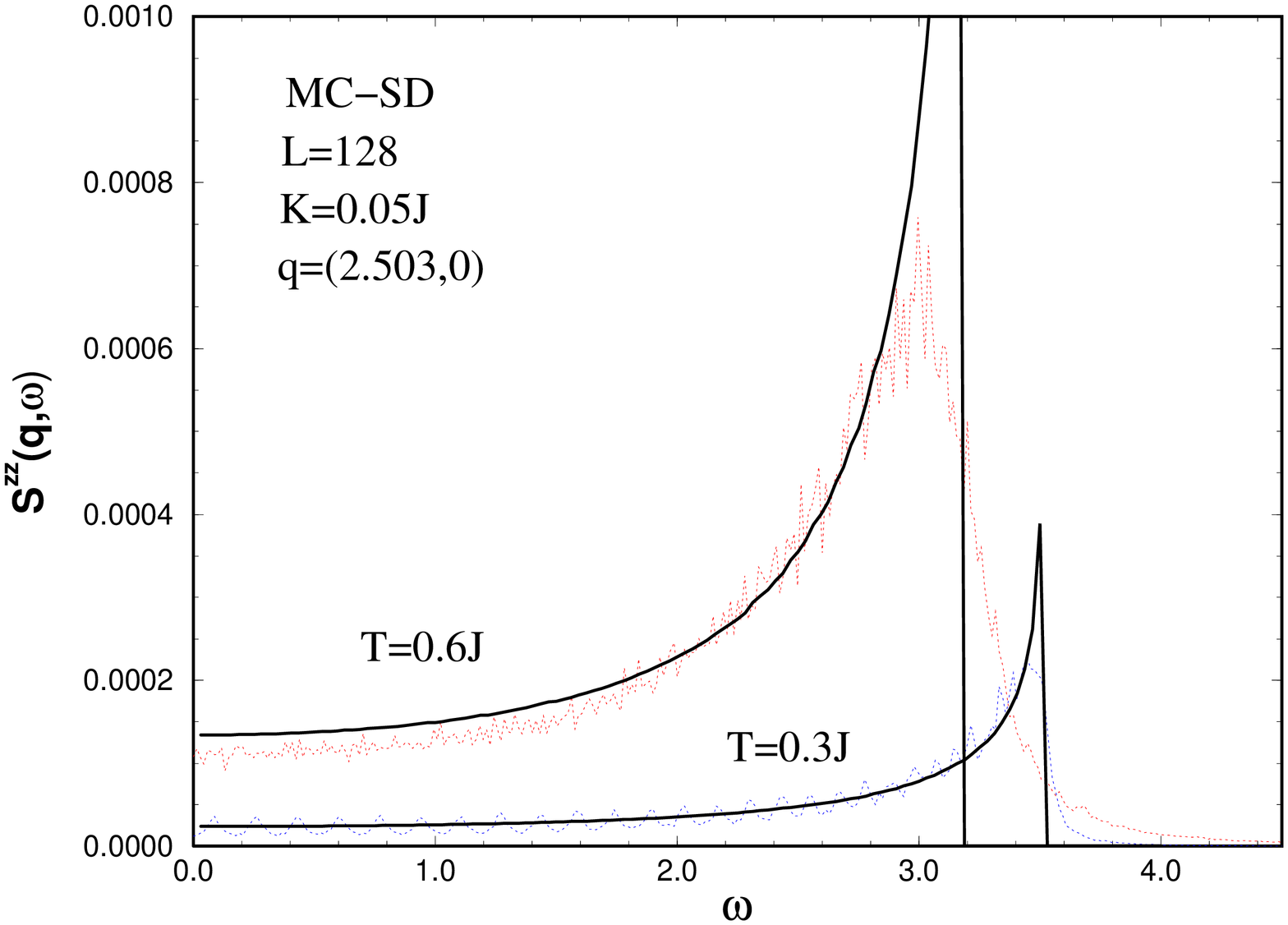,width=3.5in,angle=-90.0}
\caption{ \label{Szz2503}  $S^{zz}({\bf q},\omega)$ obtained from
(continuous line) continuum limit (\protect\ref{Szz})  and from (circles and
triangles) numerical simulation
for $K=0.05J$, $T=0.3J$, and $T=0.6J$, and
wavevector ${\bf q}=(2.503,0)$.}
\end{figure}

\begin{figure}
\psfig{file=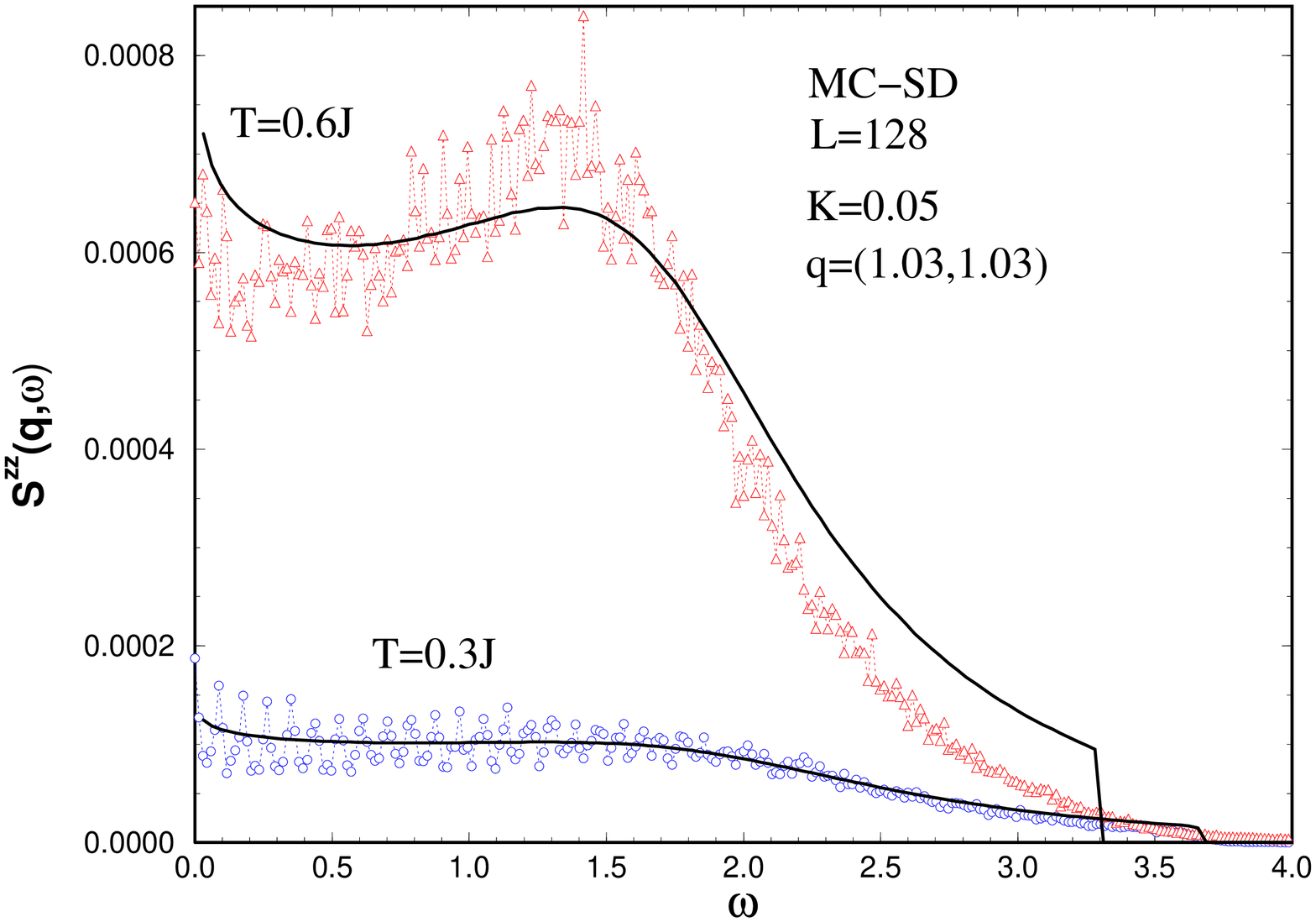,width=3.5in,angle=-90.0}
\caption{ \label{Szzd103}  $S^{zz}({\bf q},\omega)$ obtained from
(continuous line) continuum limit (\protect\ref{Szz})  and from (circles and
triangles) numerical simulation
for $K=0.05J$, $T=0.3J$, and $T=0.6J$, and
wavevector ${\bf q}=(1.03,1.03)$.}
\end{figure}

For $S^{zz}({\bf q}, \omega)$, the widths of the multiple peaks 
are determined both by the intrinsic width due to temperature, 
and the width $2\pi/t_{max}$ inherent in the spin-dynamics simulation.
For larger lattices, or higher temperatures, the spacing of the 
multiple peaks in $S^{zz}({\bf q}, \omega)$ becomes smaller than 
their measured widths, the peaks merge and the curve is much smoother.
Thus in our simulations the finite-size effects are quite well
smoothed out for lattices $L > 128$ and/or for high temperatures.
In Figures \ref{Szz0393}-\ref{Szzd103}, for $K=0.05J$, several $\bf{q}$ values,
and $L=128$ we see that the simulation data for the higher
temperature $T=0.6J$ are smooth while the data for $T=0.3J$
still show sharp peaks.

\begin{figure}
\psfig{file=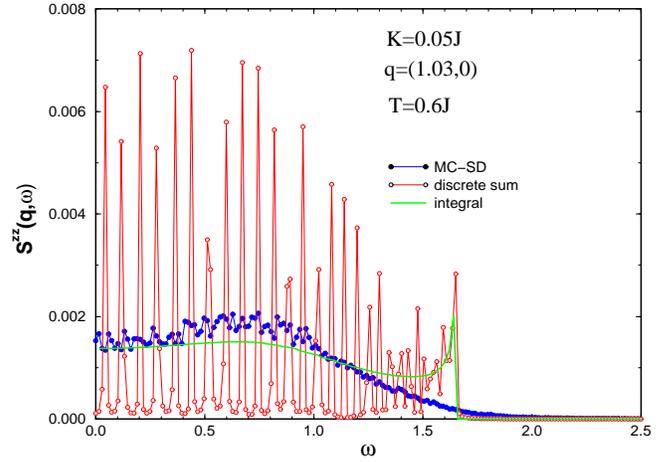,width=3.5in,angle=-90.0}
\caption{ \label{Szz3calc} $S^{zz}({\bf q},\omega)$ obtained from
numerical simulation (filled circles), discrete summation (empty circles)
and from continuum limit (line) for $K=0.05J$, $T=0.6J$, $L=128$, and
wavevector ${\bf q}=(1.03,0)$. }
\end{figure}

Using the ``discrete'' equation (\ref{disc}) for obtaining 
$S^{zz}({\bf q},\omega)$ for lattice size $L=128$ 
 we do not get 
rid of the multiple peak structure  even for $T=0.6J$.
This can be seen in Fig.\ \ref{Szz3calc} where the three
types of calculations ---
numerical simulation, discrete summation (\ref{disc})
and continuum limit (\ref{Szz}) ---
we used  to obtain $S^{zz}({\bf q},\omega)$
are shown for $K=0.05$, $T=0.6J$
and $\bf{q}=(1.03,0)$.
Typically, the discrete SCHA summation results in an 
$S^{zz}({\bf q},\omega)$ curve with very strong multiple
peak structure.
In order to smooth out the structure obtained from (\ref{disc})
it is necessary to consider much larger lattices ($L>500$).
It is natural to expect that it is more difficult to smooth out the 
spectra obtained by (\ref{disc}) than the one obtained via 
spin dynamic simulation.
%
Clearly the MC-SD calculation contains more fluctuations
and therefore greater peak widths, especially as $T$ approaches $T_c$,
whereas in expression (\ref{disc}) all spinwave peaks have very
narrow widths determined only by the integration time. 
Instead of trying to smooth the SCHA spectra by considering larger and 
larger lattices for the calculation of (\ref{disc}) --- which requires 
extra computational effort --- we can go to the continuum 
approximation limit built in (\ref{Szz}).
In fact, most real systems contain a large number $N$ of spins
($N \rightarrow \infty$) and effects due to the discreteness of the
lattice are not important.
These macroscopic systems will be better represented by the continuous
approximation built in (\ref{Szz}).
Figures \ref{Szz0393}-\ref{Szzd103} show the spectral functions obtained by
numerically evaluating (\ref{Szz}) for $K=0.05$, $T=0.3J$, and $T= 0.6J$,
for the following wavevectors:
${\bf q}=(0.393,0)$, $(1.473,0)$, $(2.50,0)$ and $(1.03,1.03)$.
These are compared with the corresponding MC-SD
calculations for $128 \times 128$ lattices.

Obviously, considering the dynamical simulation,
it is not possible to go to the $N\rightarrow\infty$ limit:
the computational cost in simulations increases tremendously
with $N$.
Nevertheless, we can remark on interesting features concerning
the results obtained from the SCHA calculation and
from numerical simulation procedures. 
First, the frequency width of the region in which 
$S^{zz}({\bf q},\omega)$ has appreciable intensity 
does not depend on the lattice size
and on the kind of calculation performed to obtain
$S^{zz}({\bf q},\omega)$.
This can be observed in
Figs.\  \ref{MC-Small}, \ref{DS-Small} and \ref{Szz0393},
which correspond to the three different ways
we have used to obtain the spectral function for different lattice
sizes but for the same wavevector ${\bf q}=(0.393,0)$. 

\begin{figure}
\psfig{file=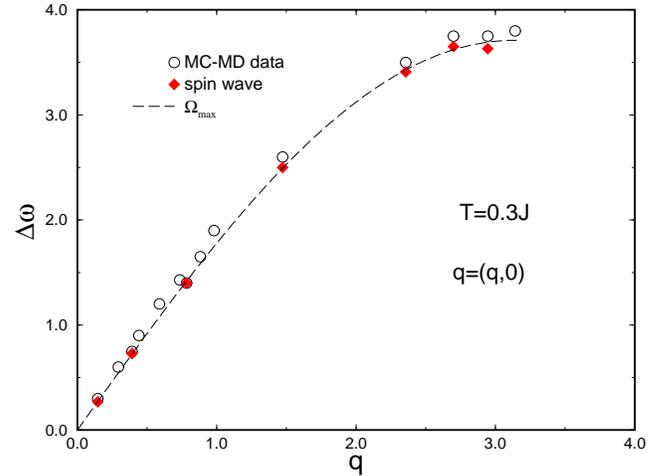,width=3.5in,angle=-90.0}
\caption{ \label{width} Width of $S^{zz}({\bf q},\omega)$ as a
function of $q$ for $K=0.05J$ and $T=0.3J$. }
\end{figure}

In Fig.\ \ref{width} we show the comparison of the width 
$\Delta\omega$ of the obtained spectral functions in 
the whole $\mid{\bf q}\mid$ range for wavevectors 
like ${\bf q}=(q,0)$: the data were obtained for
$K=0.05J$ and $T=0.3J$.
The comparison is remarkably good (a similar agreement is
obtained for $T=0.6J$).
We see that, for small $\mid{\bf q}\mid$, the width $\Delta \omega$
increases linearly with the wavevector.
A trivial analysis of (\ref{Szz}) leads us to the conclusion 
that $\Delta \omega$ must be related to the maximum 
value $\Omega$ can have for each ${\bf q}$.
From (\ref{Omega}) we easily obtain that 
$\Omega_{max} = B(T) \sin \mid q \mid /2$ where  
$B(T) = \epsilon JS[1-\beta(T)+\eta(T)(1+K)]$
and $\epsilon =1$ for  ${\bf q}=(q,0)$ wavevectors
and $\epsilon =2$ for ${\bf q}=(q_x,q_x)$.
For comparison, we show, in Fig.\ \ref{width}
a curve (dashed line) corresponding to $\Omega_{max}$.

\begin{figure}
\psfig{file=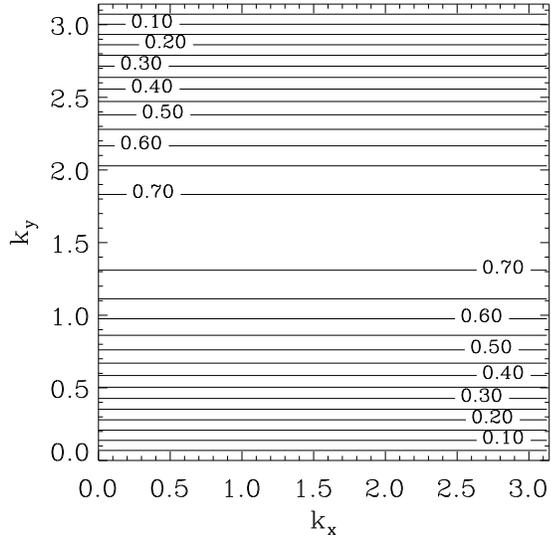,width=3.5in,angle=90.0}
\caption { \label{Omegay} Contours of difference frequency 
$\Omega(\bf{q},\bf{k})$ for $T=0.3J$ and ${\bf q}=(0,0.393)$ 
as function of $\bf{k}$. }
\end{figure}

\begin{figure}
\psfig{file=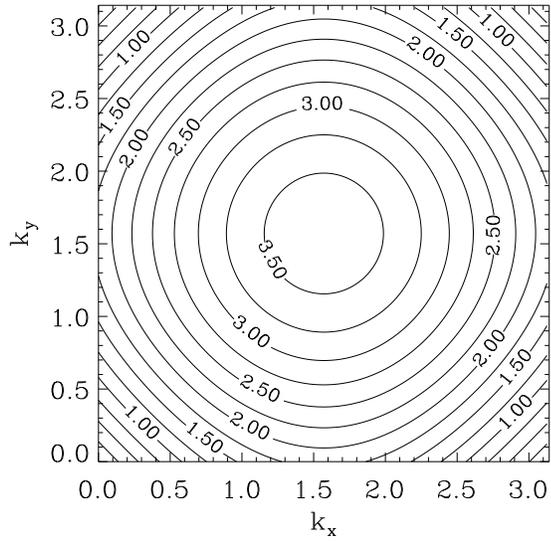,width=3.5in,angle=90.0}
\caption { \label{Omegad} Contours of difference frequency
$\Omega(\bf{q},\bf{k})$ for $T=0.3J$ and ${\bf q}=(1.03,1.03)$ 
as function of $\bf{k}$. }
\end{figure}

Second, the SCHA curves corresponding to wavevectors of the form
${\bf q}=(q,0)$ or $(0,q)$ ( Figures \ref{Szz0393}-\ref{Szz2503} ) 
show a sharp peak at higher frequencies, just before
the spectral function vanishes. 
For wavevectors like ${\bf q}=(q,q)$, this sharp peak is only observed 
near the point $(\pi, \pi)$, and not for smaller $\vert {\bf q} \vert$, 
as can be seen in Fig.\ \ref{Szzd103}.
The appearance or not of these peaks in the SCHA calculation
depends on the behavior of the density of states
$\mid\nabla\Omega\mid^{-1}$ in Eq.\ (\ref{contour}).
Figures \ref{Omegay} and \ref{Omegad} show contours of
$\Omega(\bf{q},\bf{k})$ in the  $\bf{k}$-plane,
for ${\bf q}=(0,0.393)$ and ${\bf q}=(1.03,1.03)$, respectively, 
for $T=0.3J$ .
For ${\bf q}$ in the (10) or (01) directions, the contours
are straight lines (Fig.\ \ref{Omegay}).
They become very widely spaced near $k_x=\pi/2$, or near $k_x=\pi/2$, 
where $\Omega$ approaches $\Omega_{max}$, and $\mid\nabla\Omega\mid^{-1}$ 
becomes very large along the whole straight contour.  
The integration along the contour in Eq.\ (\ref{contour})
then leads to the sharp peaks at $\omega\rightarrow \Omega_{max}$
seen in $S^{zz}({\bf q},\omega)$.
For ${\bf q}$ along the (11) direction, the contours are 
curves (Fig.\ \ref{Omegad}).
For moderate values of $\mid {\bf q} \mid$, the higher contours
(near $\Omega_{max}$) approximate small circles, having limited
total length and thus creating no sharp peak in  
$S^{zz}({\bf q},\omega)$.  
Only for ${\bf q}$ very close to the point $(\pi,\pi)$ is 
the effect due to the divergence of $\mid\nabla\Omega\mid^{-1}$
more important than the contour length, and there a
sharp peak at $\omega\rightarrow \Omega_{max}$
does occur.

It is interesting to notice that this peak is also
present in the data obtained from the discrete spin wave calculation
(Figure \ref{Szz3calc}) although the density of states
does not appear explicitly in (\ref{disc}). 
Nevertheless, the two spin wave calculations must, in fact, 
give similar results because (\ref{Szz}) or (\ref{contour})
correspond to the $t_{max}\rightarrow \infty$, and
$N\rightarrow \infty$ limits of (\ref{disc}).
For small wavevectors, this sharp high frequency peak
is not seen in the simulation data suggesting that
inclusion of higher order terms in the spin wave theory
would  probably lead to the attenuation of this peak
in the SCHA results.
As the wavevector $\mid{\bf q}\mid$ increases,
a lateral shoulder develops in the spectra obtained
from both numerical simulation and SCHA calculations;
it is already well defined for $q \sim 0.50$.
For very large wavevectors, as in  Figure \ref{Szz2503},
the lateral shoulder for the MC-SD data occurs in the
frequency region affected by the increase of 
$\mid\nabla\Omega\mid^{-1}$.
This shoulder seems to be a characteristic of 
two-spin-wave processes because it has been observed 
in other systems\cite{bethtonico}.

As the temperature increases, the width of the spectral function
$S^{zz}({\bf q},\omega)$ decreases but its height increases.
The spin wave calculation seems to agree well with the MC-SD data
for large wavevectors,  even for $T=0.6J$. 
For small wavevectors and higher temperature, however, 
$S^{zz}({\bf q},\omega)$ from the SCHA 
calculation is smaller than the MC-SD data (Fig.\ \ref{Szz0393}), 
suggesting that at high temperatures other processes could be 
contributing to the dynamical properties of this system. 
For systems with easy-axis anisotropy one can expect the formation of domains,
as in the two-dimensional Ising model, and, also, localized solitons\cite{Kovalev}.
In particular, it is usually expected\cite{Mertens} that localized solitons
would contribute to the dynamical correlation function 
in the $\omega\rightarrow 0$ (central peak) region and, mostly,
for small wavevectors.
%
%

\section{CONCLUSIONS}
\label{Conclude}

We have applied a self-consistent harmonic approximation
to the easy-axis model, obtaining the spinwave energies,
critical temperature and dynamic correlation functions.
We also demonstrated how it is possible to apply the Wolff
cluster Monte Carlo scheme to this easy-axis model, by having it act
on only the $S^z$ spin components.  
For the critical temperature, the SCHA and MC results
agree favorably over a wide range of easy-axis anisotropy,
both giving $T_c$ increasing linearly with $K$ for $K \gg J$.
The spin-dynamics calculation of dynamic correlation
functions shows interesting multiple-peak features in 
$S^{zz}({\bf q}, \omega)$, that are most easily seen in small
lattices.
These finite-size dynamical features are correctly described by 
the SCHA, especially for $T$ far below $T_c$.
Similar features should appear in models with other symmetries:
there are strong evidences that these effects were also observed
in other simulations of two dimensional easy-plane models.\cite{Evertz}

All the dynamical calculations discussed in this work were performed 
for anisotropy parameter $K=0.05$, which corresponds to a transition 
temperature $T_c=0.75J$.
For this anisotropy, two temperatures were analyzed: $T=0.3J \ll T_c$,
and $T=0.6J$.
We could not expect that the spinwave
calculation performed here, which neglects higher order terms in the spin
interactions, would reproduce exactly the simulation data.
However, the agreement for the lowest temperature, $T=0.3J$, is very good.
It is also surprisingly good for $T=0.6J$, a relatively high temperature,
and large wavevectors where a lateral peak is seen to develop.
At $T=0.6J$ , for small wavevectors and small frequencies,
the SCHA function for $S^{zz}$ shows a central peak with height
{\em smaller} than the one obtained from MC-SD simulation.
On the other hand, the SCHA prediction for the integrated intensity 
$I^{xx}$ for the {\em in-plane} correlations lies {\em above} the MC-SD
data for $T=0.6J$.
These features may suggest that other excitations, like
localized solitons and domain walls, may contribute to the dynamical 
correlation function as the temperature approaches the critical temperature.
It was shown\cite{Costa94} that the density of these localized
solitons increases exponentially  with $T$ as $T\rightarrow T_c$
and, then, one should expect that their contribution to
the dynamics of the system becomes more important
for temperatures $T \sim T_c$.
To stress this conclusion, we remark that $S^{xx}({\bf q},\omega)$
obtained by MC-SD simulation for $T>T_c$ (not shown here
because SCHA cannot be compared in this temperature regime)
does show a central peak ($\omega \sim 0$) that increases with $T$.
Its properties will be analyzed in a future work.

We conclude by saying that the two-spin wave calculation can
explain the main features obtained from Monte-Carlo-spin-dynamics 
simulation at very low temperatures.
As $T \rightarrow T_c$, the comparison between SCHA
spin-wave calculation and the numerical simulation data suggests
that other excitations may contribute to the dynamic properties of the model.
However, a better understanding concerning the contributions these
excitations might give to the dynamic spectral functions requires 
some theory which takes into account the existence of such objects.
To our knowledge, such theory for easy-axis anisotropy two-dimensional
systems is not available in the literature.

\medskip
{\sl Acknowledgements.}---The authors gratefully acknowledge the
the support of:  NSF DMR-9412300, NSF INT-9502781, 
NSF CDA-9724289, FAPEMIG,
CAPES-DAAD, and CNPq.

\newpage

\end{document}